\documentclass[12pt]{article}
\usepackage{times}
\usepackage{amsmath,amsthm,amssymb,setspace,enumitem,epsfig,titlesec,verbatim,color,array,eurosym,multirow,blkarray,lscape,soul,graphicx,cite,comment,nicefrac,lineno,epstopdf,makecell,url}
\usepackage[bf,small]{caption}
\usepackage[top=2.5cm,left=2.8cm,right=2.8cm,bottom=3.4cm]{geometry}
\DeclareMathSizes{11.5}{20}{14}{10}   
\usepackage[super,sort&compress,comma]{natbib}

\begin{document}

\onehalfspacing

\title{Embedding Sustainability in Undergraduate Mathematics with Actionable Case Studies}
\author{Maria Kleshnina$^{1,2,3,*}$, Matthew H. Holden$^{4,5,6*}$, \\ Ryan F. Heneghan$^{7,8}$, Kate J. Helmstedt$^{1,2,3,9}$\\
{\small $^1$ School of Mathematical Sciences, Queensland University of Technology, Brisbane, Australia} \\
{\small$^2$ Securing Antarctica's Environmental Future, Queensland University of Technology, Brisbane, Australia }\\
{\small$^3$ Centre for Environment and Society, Queensland University of Technology, Brisbane, Australia }\\
{\small$^4$ School of Mathematics and Physics, University of Queensland, St Lucia, Australia }\\
{\small$^5$ Centre for Biodiversity and Conservation Science, University of Queensland, St Lucia, Australia }\\
{\small$^6$ Centre for Marine Science, University of Queensland, St Lucia, Australia }\\
{\small$^7$ School of Environment and Science, Griffith University, Nathan, Australia }\\
{\small$^8$Australian Rivers Institute, Griffith University, Nathan, Australia}\\
{\small$^9$ Centre for Data Science, Queensland University of Technology, Brisbane, Australia }\\
{\small$^{*}$ these authors contributed equally}}
\date{}


\maketitle

\begin{abstract}
There is a growing need to integrate sustainability into tertiary mathematics education given the urgency of addressing global environmental challenges. This paper presents four case studies from Australian university courses that incorporate ecological and environmentally-conscious concepts into the mathematics curriculum. These case studies cover topics such as population dynamics, sustainable fisheries management, statistical inference for endangered species assessment, and mathematical modelling of climate effects on marine ecosystems. Each case demonstrates how fundamental mathematical methods, including calculus, statistics and operations research, can be applied to real-world ecological issues. These examples are ready-to-implement problems for integrating ecological thinking into mathematics classes, providing educators with practical tools to help students develop interdisciplinary problem-solving skills and prepare for the challenges of sustainability in their future careers.
\end{abstract}

\section*{Introduction}

Global environmental change demands urgent, interdisciplinary solutions as our society is faced with the need to respond to biodiversity loss, climate change, and ecosystem degradation. As stated in the United Nations Sustainable Development Goals (SDGs), education must play a key role in equipping future generations with the knowledge to address these challenges \cite{barwell2018some}. However, mathematics is often underutilised in sustainability education, and sustainability concepts are rarely integrated into mathematics curricula, despite mathematics providing the quantitative reasoning essential for informed environmental decision-making.

This gap is increasingly recognised in both mathematics and ecology education literature \cite{barwell2022mathematics, coles2024socio, karjanto2023mathematical} as well as in government policy \cite{national2015}. While a few instances document the successes of individual ecological examples in enriching the teaching of mathematical modelling \cite{vanderhoff2017interdisciplinary, koss2011sustainability, watson2025using}, the integration of environmentally-conscious topics across a broad spectrum of mathematics courses and programs remains limited. More systematic efforts are needed to normalise ecological and environmental problems as core approaches for teaching fundamental mathematical methods.

Efforts to embed quantitative skills in biology curricula mirror concerns around the poor integration of mathematics in ecological contexts. Thompson et al. \cite{thompson2013infusing} make a compelling case for the infusion of mathematical and statistical tools across the biological sciences. The authors argue that meaningful ecological understanding increasingly depends on mathematical aptitude. However, for such interdisciplinary integration to succeed, mathematics education must also evolve, adapting its examples, methods, and pedagogies to reflect the realities of socio-economic challenges and SDGs.

In this paper, we present four case studies from Australian university courses that integrate ecological and environmental concepts into the mathematics curriculum. The cases cover fundamental mathematical domains and illustrate that integrating sustainability is both feasible and valuable across subfields. The ecological topics explored include population dynamics, sustainable fisheries management, statistical inference for endangered species, and mathematical modelling of climate impacts on ecosystems. Each case study is grounded in a real-world environmental problem, employing mathematical methods that not only foster disciplinary learning but also cultivate interdisciplinary insight.

Our approach draws on recent calls to make mathematics education more applied and socially relevant through project-based and scenario-driven learning \cite{fulford2024mathematical, winkel2024using}. These case studies align with emerging frameworks for embedding sustainability in mathematics education \cite{moreno2022training, vasquez2023integrating} and reflect a pedagogical philosophy where mathematics is not just an abstract tool but a language for engaging with the living world. As shown by Winkel \cite{winkel2012sourcing} and others, mathematical models can be a powerful method for students to understand both natural systems and human impacts on those systems.

While not exhaustive, the four cases serve as adaptable examples for integrating sustainability into widely taught areas of undergraduate mathematics. Accordingly, we focus on four core topics common to most undergraduate programmes: calculus and ordinary differential equations, statistics, optimisation and operations research, and partial differential equations. The challenge here is in formulating real ecological problems to fit these existing mathematical topics. As a result, we focus on the setup of each problem, and how it links to undergraduate topics in mathematics rather than worked solutions. In the following sections, we describe each case in detail, outlining its educational objectives, mathematical structure, and ecological framing. By offering these examples, we aim to provide educators with a practical framework for embedding sustainability into mathematics courses, empowering students to become expert problem-solvers and informed decision-makers in the face of global environmental challenges.

\section*{Case studies for teaching mathematics and sustainability}

We report four case studies across four fundamental mathematical subfields as summarised in Table~\ref{tab:case-studies}.
In each example, students develop their understanding of a mathematical concept through engaging with real-world ecological challenges, which highlights the interdisciplinary nature of solving sustainability problems. These cases provide a structured approach for introducing students to the full modelling cycle: starting with simple models and adding ecological and, therefore, mathematical complexities. This mirrors scientific thinking where models evolve in response to new evidence. 

The cases are designed to support a variety of learning modes as some students excel in analytical reasoning, while others find more value in numerical analysis. Our case studies stimulate active learning of mathematics through a mix of analytic solutions and numerical methods. Each case also promotes a deeper understanding of how mathematical models influence policy and decision-making, especially in sustainability applications.

\begin{table}[h]
\begin{tabular}{|p{0.3cm}|p{3cm}|p{5cm}|p{5.5cm}|}
\hline
\textbf{\#} & \centering \textbf{Course} & \centering \textbf{Topic} & {\textbf{\centering Mathematical method} } \\
\hline
1 & \centering Calculus & Population dynamics and sustainability & Analytical and numerical solutions of ODEs \\
\hline
2 & \centering Operations Research & Allocating sustainable fisheries catch spatially & Linear programming \\
\hline
3 & \centering Statistics & Endangered species assessment & Maximum Likelihood, Bayesian parameter estimation \\
\hline
4 & \centering Partial differential equations & Climate and fishing impacts on marine ecosystems & Analytical and numerical solutions of PDEs \\
\hline
\end{tabular}
\caption{Summary of case studies integrating sustainability into tertiary mathematics education.}
\label{tab:case-studies}
\end{table}

\subsection*{Case Study 1: Population dynamics with ordinary differential equations }

In line with our approach of starting with simple models and gradually building complexity, we begin our portfolio of case studies with the fundamental question of how to model species population growth. Our focus is on understanding how species interact with their environments and how environmental factors influence population dynamics \cite{levin1989ecology}. Population dynamics models are widely used to inform policy decisions, such as managing endangered species or regulating resource extraction \cite{levin1999towards}. By applying mathematical models like the logistic growth model, students gain insight into how ecological systems evolve under different conditions, and how small changes in parameters can lead to significantly different outcomes. 

\subsubsection*{Problem Description}
The task focuses on modelling the population dynamics of some species. Initially, students work with the classic logistic growth model, which describes the population growth of the species constrained by environmental carrying capacity \cite{hallam2012mathematical}. This is represented by the logistic differential equation:
\begin{equation*}
\frac{du(t)}{dt} = r u(t)\left(1 - \frac{u(t)}{K}\right),
\end{equation*}
where $u(t)$ represents the biomass of the population at time $t$, $r$ is the intrinsic growth rate, and $K$ is the carrying capacity of the environment.

As part of the problem, students are then asked to modify this basic model by considering real-world ecological factors such as harvesting and the Allee effect \cite{stephens1999allee}. The equations to model these scenarios are:

1. Harvesting Pressure:
   \begin{equation*}
   \frac{du(t)}{dt} = r u(t) \left(1 - \frac{u(t)}{K}\right) - \delta u(t),
   \end{equation*}
   where $\delta$ is the rate of harvesting.
   
2. Allee Effect:
   \begin{equation*}
   \frac{du(t)}{dt} = r u(t) \left(\frac{u(t)}{T} - 1 \right) \left(1 - \frac{u(t)}{K} \right),
   \end{equation*}
   where $T$ represents the threshold, below which the population growth rate declines.

Students will:
\begin{itemize}
    \item Describe how parameters $r$, $K$, $\delta$, and $T$ influence the population dynamics.
    \item Analyse the fixed points of the equations and their stability.
    \item Solve the equations analytically/numerically to understand different population scenarios.
    \item Visualise the solution curves for different parameter values.
\end{itemize}

This example involves several core mathematical techniques:
\begin{itemize}
    \item \textit{Ordinary Differential Equations (ODEs):} The logistic model and its extensions are governed by first-order nonlinear differential equations. Students need to analyse the system qualitatively (fixed points, stability) and quantitatively (solving for solutions).
    \item \textit{Qualitative Analysis:} Students perform stability analysis to identify stable and unstable fixed points and use phase lines and direction fields to interpret the behaviour of the system.
    \item \textit{Numerical Methods:} For complex models like the Allee effect or harvesting, students will often need to use numerical integration methods (such as Euler’s method or Runge-Kutta methods \cite{akinsola2023numerical}) to approximate solutions when analytical solutions are not possible.
\end{itemize}

This can be adapted for a lecture by having students follow along with live demonstrations using MATLAB or Python to perform stability analysis and simulate the population dynamics under different scenarios. In an assessment setting, students could be tasked with deriving the equations, performing the analysis, and interpreting the ecological implications of their results.\\

\textbf{Connection between ecological context and mathematical approach.}
Ecology provides a rich context for teaching differential equations because it offers intuitive, real-world applications where students can visualise the effects of mathematical changes in ecological settings. Nonlinearities, thresholds, and multiple equilibria present in ecological models make them ideal for learning qualitative techniques such as stability analysis and phase portraits. However, the logistic growth model in this example simplifies ecological dynamics by focusing on a single species, overlooking complex interactions within ecosystems \cite{mckerral2023empirical}. While two-species, Lotka-Voltera, predator-prey and competition ODE systems are the standard extension to the logistic equation, they are perhaps one of only a few limited examples of ecology that have ubiquitously made it into undergraduate mathematics classrooms \cite{boyce2017elementary,strogatz2024nonlinear}, and so we avoid reviewing them here.

Instead, we would like to highlight less commonly taught models that address sustainability challenges using similar mathematical techniques. For example, models of resource extraction, such as hunting or fishing, can be framed as predator–prey systems where harvesters seek profits from removing individuals from a population. As population density declines, harvesters must exert more effort to locate individuals, introducing feedbacks that can stabilise or destabilise the population. If the price paid per unit harvested is fixed, the resulting dynamics mirror a classic predator–prey system \citep{clark1974mathematical,auger2010effects,holden2017high}, with harvesters as predators and the target species as prey. However, when price varies depending on market saturation or scarcity the dynamics become more complex. These models remain analytically tractable and offer compelling opportunities to introduce advanced concepts such as linear stability analysis, bifurcations, stable and unstable manifolds, basins of attraction, and Lyapunov functions \citep{auger2010effects,holden2017high}. For instance, students can explore whether legalising trade in animal products can, under certain conditions, reduce poaching and improve conservation outcomes, a controversial question in sustainability science \citep{holden2021poacher}. Engaging students with such models allows them to explore real-world debates and prepares them to analyse pressing questions in natural resource management. One such application, optimising sustainable catch, is the focus of the next case study, as we move from teaching differential equations to operations research and optimisation theory \cite{lenhart2007optimal, acquaye2025operational}.

\subsection*{Case Study 2: Operations Research for Sustainable Fisheries Management}

Building on the concept of population growth and resource extraction models, the next case study shifts from theoretical population dynamics to the practical application of optimisation techniques in managing natural resources.  We explore the complex issue of sustainable fisheries management, where trade-offs must be balanced between ecological health, economic outcomes, and societal needs  \cite{castonguay2023moo, erm2023biodiversity, takashina2024understanding}.  There is a rich history of using mathematical models to describe, understand, and manage harvested fish populations (i.e. fisheries)  \cite{yoshioka2024optimal, bang2024beyond}. We build on that history in the following question, where students act as mathematical advisers helping optimise a sustainable fishing license policy to learn foundational operations research theory.

\subsubsection*{Problem Description}
Students are tasked with optimising the allocation of fishing licenses for Barramundi across $n$ distinct locations. Fish are a natural resource, and so like with many natural resources the state government wants to spread the economic benefits around as much as possible. Each company out of $m$ has different costs associated with harvesting fish, and the state government wants to allocate licenses equally to ensure sustainability and minimise costs. Fish are a renewable resource, but only if their populations stay large enough to reproduce. Each fish species, depending on their biology, has a maximum sustainable yield (MSY) -- this is the amount that can be removed from each population sustainably \cite{gillis2024maximum}. Barramundi has a different MSY at each location based on the quality of the habitat and other competitors in the local ecosystem. Students are supplied two datasets: the fish population and MSY proportion at each fishing location, and the heterogeneous cost each company incurs by fishing at each location. The problem is framed as a balanced transportation problem, where the goal is to minimise the total fishing cost subject to sustainability constraints. The total sustainable supply of fish is dictated by the MSY and population at each site, and the problem setup assumes the entire supply is allocated as equal demand between the fishing companies.

For the transportation problem, students may be asked to formulate the full linear programming formulation as above, or use transportation-problem-specific shortcuts to more quickly solve the problem with the transportation simplex method. In the transportation simplex method, the supply nodes are the fishing locations, and each company is a demand node. Each company requires an equal share of the maximum sustainable yield. The full linear programming mathematical formulation of the problem is:
\begin{equation*}
\min \sum_{i=1}^{n} \sum_{j=1}^{m} c_{ij} x_{ij},
\end{equation*}
where $c_{ij}$ is the cost of harvesting from location $i$ by company $j$, and $x_{ij}$ is the number of tonnes allocated to company $j$ at location $i$, subject to:
\begin{equation*}
\sum_{j=1}^{m} x_{ij} = MSY_i, \quad \forall i,
\end{equation*}
ensuring that the total allocation for each location does not exceed the MSY, and demand constraints ensuring each company receives the same yield and the entire MSY is distributed.

Students will:
\begin{itemize}
    \item Learn how to formulate the problem using linear programming.
    \item Apply methods like the North West Corner Method or Vogel's Approximation for finding an initial solution.
    \item Use the transportation simplex method (or the full simplex method) to find the optimal solution.
    \item Make a clear, declarative recommendation of the amount each company should fish at each site.
\end{itemize}

In mathematical techniques, this case involves:
\begin{itemize}
    \item \textit{Linear Programming (LP):} The problem is formulated as a transportation problem, a classic LP problem where students minimise the total cost of fishing, subject to constraints on sustainability (MSY).
    \item \textit{Optimisation Methods:} Students use the simplex method to solve the LP problem.
    \item \textit{Sustainability Constraints:} The MSY constraint ensures that the allocation respects ecological limits, demonstrating the intersection of optimisation and environmental sustainability.
\end{itemize}

In a lecture the instructor could walk through the formulation of the problem and demonstrate the relationship between the simplex method the students already understand and the requirements, assumptions, and setup for the transportation simplex method. In an assessment, students could be tasked with converting the word problem to an appropriate LP formulation, analysing the problem in depth, and making formal recommendations based on their results.\\

\textbf{Connection between ecological context and mathematical approach.}
Operations research offers powerful analytical tools for making complex decisions. These tools are useful across a broad range of biodiversity conservation and sustainability challenges, where limited resources must be allocated under uncertainty to achieve competing ecological, societal, and economic objectives. Biodiversity conservation is focused on protecting species, ecosystems, and habitats in perpetuity \cite{hemming2022introduction}, taking actions like placing protected areas, restoring habitat, or performing recovery or supportive actions on particular species or populations.  Many of these actions naturally map to classical OR problem classes including integer and mixed-integer programming (specifically when asking where actions should be performed in space \cite{albers2024anticipating, drechsler2017costs,armsworth2025multilevel}, or when prioritising which species to support \cite{helmstedt2016prioritizing}), multi-objective optimisation (especially in areas that support multiple species or have economic uses \cite{cho2023understanding}), and dynamic programming (when we must act many times in succession \cite{helmstedt2017costs}). These real-world applications can be simplified, or in their full, complex form can feature tangible constraints, nonlinear trade-offs, and spatio-temporal dynamics, making them ideal case studies for teaching foundational or advanced OR techniques. Beyond biodiversity conservation, managing environmental resources sustainably - particularly agriculture \cite{pla2014perspective}, fisheries \cite{bjorndal2004operational}, or renewable energy provision \cite{thomas2024operational} -  requires many similar tradeoffs and complexities, providing ample opportunities for OR.

This case study can be extended to incorporate the uncertainty in both the location and population of fish, which is a common challenge in real-world fisheries management. In many cases, we may not know exactly where fish are located or the true size of the population in each area due to factors like fish migration, environmental variability, and incomplete data. One approach to model this uncertainty can incorporate stochastic elements into the problem. For instance, instead of assuming that the fish population in each location is fixed, we could model it probabilistically. This could involve treating the population sizes as random variables with distributions based on historical data, or incorporating a Bayesian framework to update beliefs about fish populations as more data becomes available. 

This brings us to the next fundamental step in ecological resource management: estimating quantities from partial observations. The following example explores statistical inference techniques, including maximum likelihood and Bayesian estimation, which help in drawing reliable conclusions from such data.

\subsection*{Case Study 3: Statistical Inference for Population Estimation}

Estimating some quantity from a subset of observations is perhaps one of the most fundamental goals of statistical inference, allowing us to draw reliable conclusions about a much larger group without examining every individual. But it's also fundamental for effective resource management and environmental decision-making. For instance, fishery scientists rely on “logbooks,” records of boats’ daily catches and effort, to infer the total number of fish in a region, rather than exhaustively counting all fish one by one \cite{hutton2019harvest}. By applying mathematical theory, statisticians transform such partial data into estimates of total abundance, complete with measures of uncertainty. These estimates are essential for guiding decisions, such as determining safe harvest levels or prioritising areas for habitat protection, and for monitoring trends over time, ensuring that management actions are based on sound evidence. The goal of the problem that follows is to provide a simple, yet compelling, introduction to the topic. Students are asked to determine if a species is threatened given field data.

\subsubsection*{Problem Description}
In this case study, students estimate the population size of a lizard species based on a subset of observations from a biodiversity survey. The task involves using maximum likelihood estimation (MLE) and Bayesian inference to estimate the total population size and assess the likelihood of the species being threatened.

The key equations include the binomial distribution for modelling the probability of observing $x$ individuals from a total population of size $n$, given a detection probability $p$:
\begin{equation*}
\text{Pr}\,(X=x\,|\,n,p) = \binom{n}{x} p^x (1-p)^{n-x},
\end{equation*}
and the corresponding likelihood function for the total population size $n$ given the data $x$ and the parameter $p$. Students will:
\begin{itemize}
    \item Derive the likelihood function, compute the MLE for $n$, and determine if the estimate is biased.
    \item Conduct a Bayesian analysis to estimate the posterior distribution for $n$.
    \item Assess whether the population size is likely to be below the threshold of 100 individuals, indicating that the species is threatened.
\end{itemize}

Here, the example relies on:
\begin{itemize}
    \item \textit{Probability Modelling:} The binomial distribution is used to model the observation process, with students deriving the likelihood function.
    \item \textit{Maximum Likelihood Estimation (MLE):} Students apply MLE to estimate the total population size.
    \item \textit{Bayesian Inference:} Students use Bayesian methods to estimate the posterior distribution for $n$, which both quantifies uncertainty in their estimate and incorporates prior knowledge about the species.
\end{itemize}

This case can be adapted for lectures by guiding students through the derivation of the likelihood function and the Bayesian update process. For assessment, students might write a report discussing their assumptions and presenting their findings. A particular strength of this problem is that the likelihood function is defined over a discrete domain, requiring students to engage more thoughtfully with the concept of maximum likelihood. Rather than following routine procedures, such as taking the derivative of the log-likelihood function and setting it equal to zero, students must reflect on the underlying logic of the method. This shift encourages deeper conceptual understanding and helps solidify the abstract principles behind statistical inference.\\

\textbf{Connection between ecological context and mathematical approach.}
Ecology provides real-world examples where statistical methods are essential for making decisions under uncertainty. Population estimation is a key issue in conservation, and statistical modelling helps students understand how to deal with incomplete data. This case study demonstrates the power of statistics in making informed decisions about the management of endangered species, showcasing the intersection of mathematics and sustainability. 

Although this problem assumes a fixed detection probability and randomly distributed identically distributed individuals in the landscape, real-world ecological surveys often violate these assumptions. In practice, both the probability of detection and the spatial distribution of individuals are typically unknown and may vary across the landscape. Estimating these quantities jointly with population size introduces additional complexity and uncertainty. These challenges provide natural extensions for more advanced courses, where students can explore hierarchical models or spatial capture–recapture methods that explicitly account for heterogeneity in detectability and occupancy \citep{royle2004n, kery2011hierarchical, chandler2013spatial, guillera2017modelling}.

Other ways to construct or extend this problem are by giving students time series data from fisher logbooks and asking them to conduct a fisheries stock assessment. This is potentially more challenging and can be done after the above example. It is the type of example where the maximum likelihood estimator may not be tractable analytically, which leads to a more computationally challenging problem. This allows the instructor to explore more advanced topics with students, such as parameter identifiability \cite{stigter2024computing}.

Building on the Bayesian approach to estimating fish populations from incomplete data, we move to modelling the dynamics of marine ecosystems using more complex mathematical methods. While Bayesian inference helps estimate population size and uncertainty, it often assumes static or simplified environmental conditions. However, marine ecosystems are dynamic and continuously affected by factors such as climate change and fishing pressure. To account for these complexities, we turn to partial differential equations (PDEs), which offer a powerful way to model how populations evolve over time and space, factoring in environmental variables like temperature, food availability, and migration patterns. The next example introduces the McKendrick-von Foerster equation, a PDE widely used to model marine ecosystems, and explores how these models can be adapted to assess the impacts of climate change and fishing pressures on species distribution and ecosystem health.

\subsection*{Case Study 4: Mathematical Modelling of Marine Ecosystems}

Marine ecosystems are under immense pressure from climate change and overfishing, both of which affect the health of these ecosystems and the services they provide. Mathematical models are crucial for understanding how different factors, such as temperature changes and fishing pressure, impact marine life. This case study introduces students to the McKendrick-von Foerster equation, a partial differential equation (PDE) used to represent marine ecosystems.

\subsubsection*{Problem Description}
Students are tasked with modelling the dynamics of marine ecosystems using the McKendrick-von Foerster equation, which describes the size-spectrum of marine organisms. The equation is given by:
\begin{equation*}
\frac{\partial N(w,t)}{\partial t} = -g(w,t) \frac{\partial N(w,t)}{\partial w} - \mu(w,t) N(w,t),
\end{equation*}
where $N(w,t)$ is the number of individuals of size $w$ at time $t$, $g(w,t)$ represents their growth rates, and $\mu(w,t)$ their mortality rates. This equation underpins numerous published, global-scale marine ecosystem models that have been used to explore the impacts of climate change and fishing on marine ecosystems \cite{blanchard_potential_2012, carozza_ecological_2015, dalaut_which_2025, heneghan_functional_2020}. The equation can be derived from biological first-principles for understanding in lectures or can be given as an example of a fully formulated partial differential equation depending on the goals of the course. For a detailed derivation of the McKendrick-von Foerster partial differential equation, see chapter 23 in \cite{kot2001elements}.

Growth and mortality rates, as well as boundary and initial conditions have been formulated in different ways across marine ecosystem models underpinned by the McKendrick-von Foerster equation. For example, growth and mortality rates can be fuelled by explicit, size-based predator-prey interactions, where larger organisms eat small \cite{blanchard_potential_2012, heneghan_functional_2020, dalaut_which_2025} or are governed by empirical scalings with body size \cite{carozza_ecological_2015}. With respect to boundary conditions, the abundance of organisms in the smallest size class can be held constant \cite{heneghan_functional_2020}, or be replenished by a flux of new recruits from reproduction of larger organisms \cite{blanchard_potential_2012,dalaut_which_2025,carozza_ecological_2015}. These rates are temperature-dependent, allowing different models to capture how rising temperatures from climate change flow through from individual-level physiology to ecosystem structure and function.

Finally, the impacts of fishing on larger size classes can be included as an additional mortality term. Similar to growth and mortality rates the representation of fishing varies across published models, from a fixed removal rate \cite{blanchard_potential_2012} to more complex formulations that link fishing pressure with the economics of fisheries \cite{carozza_formulation_2017}.

Students will:
\begin{itemize}
    \item Derive the governing equation for a marine ecosystem model from biological principles. 
    \item Explore the process of model development, from defining a research question, identifying and justifying assumptions and limitations of a chosen modelling approach within the scope of the research question, and translating biological and ecological processes into a mathematical formulation.
    \item Analyse the impact of temperature changes and fishing on the structure and function of marine ecosystems. Students will gain experience in using mathematics to develop new insights into natural systems and how humans affect them.
    \item Perform numerical simulations to explore different scenarios (e.g., varying temperature sensitivities, fishing pressures) or model formulations (e.g., using predator-prey interactions for growth and mortality versus empirical scalings with body size).
\end{itemize}

Mathematically, the students will be exposed to:
\begin{itemize}
    \item \textit{Partial Differential Equations (PDEs):} The McKendrick-von Foerster equation is a transport equation, a common type of PDE taught in university PDE or multivariable calculus courses. 
    \item \textit{Numerical Methods:} Students will use finite difference methods or other numerical techniques to approximate solutions to the PDE.
    \item \textit{Sensitivity Analysis:} Students will analyse how different parameters (e.g., fishing pressure, temperature) or model structure (e.g., changing boundary conditions) affect the modelled ecosystem over time.
\end{itemize}

For lectures, the instructor could develop a marine ecosystem model governed by the McKendrick-von Foerster equation from biological first principles. This process would be critical for students to grasp the ecological basis for the model. These lectures could be broadened by beginning with the development of a simpler marine ecosystem model that represents marine ecosystems by body size, but using only power-law relationships\cite{jennings_global-scale_2008}. This simpler model has also been used to explore the impacts of climate change on marine ecosystems \cite{tittensor_next-generation_2021}, but does not require a deep knowledge of calculus or numerical methods.
In workshops, the instructor could then guide students through the process of solving the model numerically (see Chapter 20 of \cite{press_numerical_2007}). For assessment, students could use the model to explore the impacts of different environmental changes and fishing practices on the health of the ecosystem.\\

\textbf{Connection between ecological context and mathematical approach.}
Ecology provides a powerful context for teaching PDEs because it connects abstract mathematical concepts to real-world environmental issues. The McKendrick-von Foerster equation is widely used to model marine ecosystems \cite{heneghan_disentangling_2021, tittensor_next-generation_2021}, and working with this model allows students to explore the effects of climate change and fishing using state-of-the-art modelling tools. 

Body size is the master trait in marine ecosystems, setting the pace of life and serving as a major determinant of an organism's position in the food chain \cite{andersen_characteristic_2016}. However, traits such as body composition, feeding behaviour or reproductive strategy are also important for determining an organism's relative fitness in a given environment. An extension to this case study would be to explore how these secondary traits can be incorporated into marine models governed by the McKendrick-von Foerster equation\cite{blanchard_bacteria_2017}. This would allow students to explore further questions around ecosystem structure, and the co-existence of different functional groups across changing environmental gradients and fishing pressures. 

At the same time, the oceans are the world's largest carbon sink, absorbing 25-30\% of human emissions annually\cite{nowicki_quantifying_2022}. The bulk of marine carbon sequestration is driven by plankton, however larger animals such as fish may serve a significant role as well \cite{bianchi_estimating_2021}. By quantifying carbon flow through marine food chains \cite{heneghan_functional_2020}, size-based marine models underpinned by the McKendrick-von Foerster equation allow students to explore how carbon sequestration across the marine size-spectrum may change in the future under warming, and how direct human impacts such as fishing or conservation planning may affect this critical ecosystem service\cite{bianchi_estimating_2021}.

\section*{Discussion}

This paper discusses four ready-to-implement examples that integrate sustainability problems into fundamental mathematical concepts, demonstrating how mathematics can be applied to our understanding of environmental challenges. Our cases span four mathematical domains, from calculus and operations research to statistics, addressing topics such as population dynamics, sustainable fisheries management, endangered species assessment, and the impacts of climate change on marine ecosystems. These examples teach students how to address environmental challenges with the aid of analytical thinking and rigorous mathematical modelling.

A unifying feature across case studies presented here is the role of data visualisation in helping students interpret and communicate mathematical models. Each case involves visualising model outputs, often in the form of simulated data, such as population trends or distributions, offering opportunities to teach core principles of graphical representation, pattern recognition, and interpretation. Whether exploring species distributions in invasive species management or analysing population dynamics in fisheries and conservation, visual tools make abstract models more intuitive and engaging. Further, such visualisation skills can be readily extended to real-world datasets, preparing students to analyse and communicate ecological patterns in applied contexts. As students work through these cases, they gain experience with visualisation as a bridge between quantitative models and decision-making, an essential skill in sustainability research and policy-making. Data visualisation can be seamlessly integrated into any mathematics course involving modelling, enriching students’ ability to engage with complex systems through both a visual and analytical lens.

To make this integration accessible, we provide curriculum-aligned problems that instructors can easily adapt for use in undergraduate mathematics courses. The challenge lies in formulating real-world ecological problems to fit the existing mathematical theory and techniques, so we focus on formulating problems that connect to four core undergraduate topics in mathematics, rather than providing worked-out solutions. These case studies encourage critical thinking by leaving solutions open for students to solve. This approach not only reinforces understanding of mathematical concepts but also prepares students to apply their knowledge outside of the classroom. The problems presented here can be used as assessments, where students apply techniques from ODEs, operations research, statistics, and PDEs to ecological contexts. The emphasis is on problem formulation and the connections to mathematical principles, which we believe are the most valuable outcomes for students as they work through complex environmental problems.

\section*{Acknowledgements}

This work was supported by the Australian Research Council Discovery Early Career Researcher Awards DE250101223 (to M.K.) and  DE250100013 (to R.F.H.); Australian Research Council Discovery Project  DP250102530 (to M.H.H.); Australian Research Council Special Research Initiative, SRIEAS Grant SR200100005, Securing Antarctica’s Environmental Future (to K.J.H. and M.K.).

\section*{Declaration of interest statement}

The authors report there are no competing interests to declare. 


\bibliographystyle{vancouver}
\bibliography{refs}

\begin{thebibliography}{10}

\bibitem{barwell2018some}
Barwell R.
\newblock Some thoughts on a mathematics education for environmental sustainability.
\newblock In: The philosophy of mathematics education today. Springer; 2018. p. 145-60.

\bibitem{barwell2022mathematics}
Barwell R, Boylan M, Coles A.
\newblock Mathematics education and the living world: A dialogic response to a global crisis.
\newblock The Journal of Mathematical Behavior. 2022;68:101013.

\bibitem{coles2024socio}
Coles A, Solares-Rojas A, le~Roux K.
\newblock Socio-ecological gestures of mathematics education.
\newblock Educational Studies in Mathematics. 2024;116(2):165-83.

\bibitem{karjanto2023mathematical}
Karjanto N.
\newblock Mathematical modeling for sustainability: How can it promote sustainable learning in mathematics education?
\newblock arXiv preprint arXiv:230713663. 2023.

\bibitem{national2015}
Committee NMS, et~al.
\newblock National marine science plan 2015-2025: driving the development of Australia's blue economy.
\newblock National Marine Science Committee; 2015.

\bibitem{vanderhoff2017interdisciplinary}
van~der Hoff Q.
\newblock Interdisciplinary education--a predator--prey model for developing a skill set in mathematics, biology and technology.
\newblock International Journal of Mathematical Education in Science and Technology. 2017;48(6):928-38.

\bibitem{koss2011sustainability}
Koss L.
\newblock Sustainability in a differential equations course: a case study of Easter Island.
\newblock International Journal of Mathematical Education in Science and Technology. 2011;42(4):545-53.

\bibitem{watson2025using}
Watson GS, Watson JA.
\newblock Using animal trails to illustrate mathematical functions: from beach bivalves to moth larvae.
\newblock International Journal of Mathematical Education in Science and Technology. 2025:1-13.
\newblock Early access.

\bibitem{thompson2013infusing}
Thompson KV, Cooke TJ, Fagan WF, Gulick D, Levy D, Nelson KC, et~al.
\newblock Infusing quantitative approaches throughout the biological sciences curriculum.
\newblock International Journal of Mathematical Education in Science and Technology. 2013;44(6):817-33.

\bibitem{fulford2024mathematical}
Fulford G.
\newblock Mathematical modelling using scenarios, case studies and projects in early undergraduate classes.
\newblock International Journal of Mathematical Education in Science and Technology. 2024;55(2):468-79.

\bibitem{winkel2024using}
Winkel B.
\newblock Using modelling to motivate and teach differential equations.
\newblock International Journal of Mathematical Education in Science and Technology. 2024;55(2):193-7.

\bibitem{moreno2022training}
Moreno-Pino FM, Jim{\'e}nez-Fontana R, Domingo JMC, Goded PA.
\newblock Training in mathematics education from a sustainability perspective: A case study of university teachers’ views.
\newblock Education Sciences. 2022;12(3):199.

\bibitem{vasquez2023integrating}
V{\'a}squez C, Alsina {\'A}, Seckel MJ, Garc{\'\i}a-Alonso I.
\newblock Integrating sustainability in mathematics education and statistics education: A systematic review.
\newblock Eurasia Journal of Mathematics, Science and Technology Education. 2023;19(11):em2357.

\bibitem{winkel2012sourcing}
Winkel BJ.
\newblock Sourcing for parameter estimation and study of logistic differential equation.
\newblock International Journal of Mathematical Education in Science and Technology. 2012;43(1):67-83.

\bibitem{levin1989ecology}
Levin SA.
\newblock Ecology in theory and application.
\newblock In: Applied Mathematical Ecology. Springer; 1989. p. 3-8.

\bibitem{levin1999towards}
Levin SA.
\newblock Towards a science of ecological management.
\newblock Conservation Ecology. 1999;3(2).

\bibitem{hallam2012mathematical}
Hallam TG, Levin SA.
\newblock Mathematical ecology: an introduction. vol.~17.
\newblock Springer Science \& Business Media; 2012.

\bibitem{stephens1999allee}
Stephens PA, Sutherland WJ, Freckleton RP.
\newblock What is the Allee effect?
\newblock Oikos. 1999:185-90.

\bibitem{akinsola2023numerical}
Akinsola V.
\newblock Numerical methods: Euler and Runge-Kutta.
\newblock In: Qualitative and Computational Aspects of Dynamical Systems. IntechOpen; 2023. .

\bibitem{mckerral2023empirical}
McKerral JC, Kleshnina M, Ejov V, Bartle L, Mitchell JG, Filar JA.
\newblock Empirical parameterisation and dynamical analysis of the allometric Rosenzweig-MacArthur equations.
\newblock Plos one. 2023;18(2):e0279838.

\bibitem{boyce2017elementary}
Boyce WE, DiPrima RC, Meade DB.
\newblock Elementary differential equations.
\newblock John Wiley \& Sons; 2017.

\bibitem{strogatz2024nonlinear}
Strogatz SH.
\newblock Nonlinear dynamics and chaos: with applications to physics, biology, chemistry, and engineering.
\newblock Chapman and Hall/CRC; 2024.

\bibitem{clark1974mathematical}
Clark CW.
\newblock Mathematical bioeconomics; 1974.

\bibitem{auger2010effects}
Auger P, Mchich R, Ra{\"\i}ssi N, Kooi BW.
\newblock Effects of market price on the dynamics of a spatial fishery model: Over-exploited fishery/traditional fishery.
\newblock Ecological complexity. 2010;7(1):13-20.

\bibitem{holden2017high}
Holden MH, McDonald-Madden E.
\newblock High prices for rare species can drive large populations extinct: the anthropogenic Allee effect revisited.
\newblock Journal of theoretical biology. 2017;429:170-80.

\bibitem{holden2021poacher}
Holden MH, Lockyer J.
\newblock Poacher-population dynamics when legal trade of naturally deceased organisms funds anti-poaching enforcement.
\newblock Journal of Theoretical Biology. 2021;517:110618.

\bibitem{lenhart2007optimal}
Lenhart S, Workman JT.
\newblock Optimal control applied to biological models.
\newblock Chapman and Hall/CRC; 2007.

\bibitem{acquaye2025operational}
Acquaye A.
\newblock Operational research for sustainability: a synthesis of methods, applications and challenges.
\newblock Journal of the Operational Research Society. 2025:1-35.

\bibitem{castonguay2023moo}
Castonguay AC, Polasky S, Holden MH, Herrero M, Chang J, Mason-D’Croz D, et~al.
\newblock MOO-GAPS: A multi-objective optimization model for global animal production and sustainability.
\newblock Journal of Cleaner Production. 2023;396:136440.

\bibitem{erm2023biodiversity}
Erm P, Balmford A, Holden MH.
\newblock The biodiversity benefits of marine protected areas in well-regulated fisheries.
\newblock Biological Conservation. 2023;284:110049.

\bibitem{takashina2024understanding}
Takashina N.
\newblock Understanding the impact of selective fishery and bycatch on stock dynamics.
\newblock Natural Resource Modeling. 2024;37(4):e12403.

\bibitem{yoshioka2024optimal}
Yoshioka H.
\newblock Optimal harvesting policy for biological resources with uncertain heterogeneity for application in fisheries management.
\newblock Natural Resource Modeling. 2024;37(2):e12394.

\bibitem{bang2024beyond}
Bang RN, Steinshamn SI.
\newblock Beyond age-structured single-species management: Optimal harvest selectivity in the face of predator--prey interactions.
\newblock Natural Resource Modeling. 2024;37(2):e12393.

\bibitem{gillis2024maximum}
Gillis DM, Koscielny J, Blanz B.
\newblock Maximum sustainable yield as a reference point in the presence of fishing effort that follows an ideal free distribution.
\newblock Natural Resource Modeling. 2024;37(1):e12390.

\bibitem{hemming2022introduction}
Hemming V, Camaclang AE, Adams MS, Burgman M, Carbeck K, Carwardine J, et~al.
\newblock An introduction to decision science for conservation.
\newblock Conservation biology. 2022;36(1):e13868.

\bibitem{albers2024anticipating}
Albers HJ, Chang CH, Dissanayake ST, Helmstedt KJ, Kroetz K, Dilkina B, et~al.
\newblock Anticipating anthropogenic threats in acquiring new protected areas.
\newblock Conservation Biology. 2024;38(2):e14176.

\bibitem{drechsler2017costs}
Drechsler M, W{\"a}tzold F.
\newblock Costs of uncoordinated site selection with multiple ecosystem services.
\newblock Natural Resource Modeling. 2017;30(1):10-29.

\bibitem{armsworth2025multilevel}
Armsworth PR, Dilkina B, Jackson HB, Kroetz K, Sims C.
\newblock Multilevel Decision-Making and Protected Area Prioritization.
\newblock Natural Resource Modeling. 2025;38(3):e70004.

\bibitem{helmstedt2016prioritizing}
Helmstedt KJ, Shaw JD, Bode M, Terauds A, Springer K, Robinson SA, et~al.
\newblock Prioritizing eradication actions on islands: it's not all or nothing.
\newblock Journal of Applied Ecology. 2016;53(3):733-41.

\bibitem{cho2023understanding}
Cho SH, Mingie JC, Kang N, Zhu G, Upendram S.
\newblock Understanding the differences between single-and multiobjective optimization for the conservation of multiple species.
\newblock Natural Resource Modeling. 2023;36(1):e12356.

\bibitem{helmstedt2017costs}
Helmstedt KJ, Possingham HP.
\newblock Costs are key when reintroducing threatened species to multiple release sites.
\newblock Animal Conservation. 2017;20(4):331-40.

\bibitem{pla2014perspective}
Pl{\`a} LM, Sandars DL, Higgins AJ.
\newblock A perspective on operational research prospects for agriculture.
\newblock Journal of the Operational Research Society. 2014;65(7):1078-89.

\bibitem{bjorndal2004operational}
Bj{\"o}rndal T, Lane DE, Weintraub A.
\newblock Operational research models and the management of fisheries and aquaculture: A review.
\newblock European Journal of Operational Research. 2004;156(3):533-40.

\bibitem{thomas2024operational}
Thomas J, Weber GW, Aguilar RR, Munapo E, Vasant P.
\newblock Operational Research for Renewable Energy and Sustainable Environments.
\newblock IGI Global; 2024.

\bibitem{hutton2019harvest}
Hutton T, O'Neill MF, Leigh GM, Holden MH, Deng RA, Plaganyi E.
\newblock Harvest strategies for the Torres Strait finfish fishery. 2019.

\bibitem{royle2004n}
Royle JA.
\newblock N-mixture models for estimating population size from spatially replicated counts.
\newblock Biometrics. 2004;60(1):108-15.

\bibitem{kery2011hierarchical}
K{\'e}ry M, Schaub M.
\newblock Hierarchical modeling and inference in ecology: the analysis of data from populations, metapopulations and communities.
\newblock Academic Press; 2011.

\bibitem{chandler2013spatial}
Chandler RB, Royle JA.
\newblock Spatially explicit models for inference about density in unmarked or partially marked populations.
\newblock The Annals of Applied Statistics. 2013;7(2):936-54.

\bibitem{guillera2017modelling}
Guillera-Arroita G.
\newblock Modelling of species distributions, range dynamics and communities under imperfect detection: advances, challenges and opportunities.
\newblock Ecography. 2017;40(2):281-95.

\bibitem{stigter2024computing}
Stigter J.
\newblock Computing parameter identifiability and other structural properties for natural resource models.
\newblock Natural Resource Modeling. 2024;37(1):e12382.

\bibitem{blanchard_potential_2012}
Blanchard JL, Jennings S, Holmes R, Harle J, Merino G, Allen JI, et~al.
\newblock Potential consequences of climate change for primary production and fish production in large marine ecosystems.
\newblock Philosophical Transactions of the Royal Society B: Biological Sciences. 2012 Nov;367(1605):2979-89.
\newblock Publisher: Royal Society.
\newblock Available from: \url{https://royalsocietypublishing.org/doi/10.1098/rstb.2012.0231}.

\bibitem{carozza_ecological_2015}
Carozza DA, Bianchi D, Galbraith ED.
\newblock The ecological module of {BOATS}-1.0: a bioenergetically-constrained model of marine upper trophic levels suitable for studies of fisheries and ocean biogeochemistry.
\newblock Geoscientific Model Development Discussions. 2015 Dec;8(12):10145-97.
\newblock Available from: \url{https://www.geosci-model-dev-discuss.net/8/10145/2015/}.

\bibitem{dalaut_which_2025}
Dalaut L, Barrier N, Lengaigne M, Rault J, Ariza A, Belharet M, et~al.
\newblock Which processes structure global pelagic ecosystems and control their trophic functioning? {Insights} from the mechanistic model {APECOSM}.
\newblock Progress in Oceanography. 2025 Jul;235:103480.
\newblock Available from: \url{https://www.sciencedirect.com/science/article/pii/S0079661125000680}.

\bibitem{heneghan_functional_2020}
Heneghan RF, Everett JD, Sykes P, Batten SD, Edwards M, Takahashi K, et~al.
\newblock A functional size-spectrum model of the global marine ecosystem that resolves zooplankton composition.
\newblock Ecological Modelling. 2020 Nov;435:109265.
\newblock Available from: \url{https://www.sciencedirect.com/science/article/pii/S0304380020303355}.

\bibitem{kot2001elements}
Kot M.
\newblock Elements of mathematical ecology.
\newblock Cambridge University Press; 2001.

\bibitem{carozza_formulation_2017}
Carozza DA, Bianchi D, Galbraith ED.
\newblock Formulation, {General} {Features} and {Global} {Calibration} of a {Bioenergetically}-{Constrained} {Fishery} {Model}.
\newblock PLOS ONE. 2017 Jan;12(1):e0169763.
\newblock Available from: \url{http://dx.plos.org/10.1371/journal.pone.0169763}.

\bibitem{jennings_global-scale_2008}
Jennings S, Mélin F, Blanchard JL, Forster RM, Dulvy NK, Wilson RW.
\newblock Global-scale predictions of community and ecosystem properties from simple ecological theory.
\newblock Proceedings of the Royal Society B: Biological Sciences. 2008 Jun;275(1641):1375-83.
\newblock Available from: \url{https://royalsocietypublishing.org/doi/10.1098/rspb.2008.0192}.

\bibitem{tittensor_next-generation_2021}
Tittensor DP, Novaglio C, Harrison CS, Heneghan RF, Barrier N, Bianchi D, et~al.
\newblock Next-generation ensemble projections reveal higher climate risks for marine ecosystems.
\newblock Nature Climate Change. 2021 Nov;11(11):973-81.
\newblock Number: 11 Publisher: Nature Publishing Group.
\newblock Available from: \url{https://www.nature.com/articles/s41558-021-01173-9}.

\bibitem{press_numerical_2007}
Press WH, Teukolsky SA, Vetterling WT, Flannery BP.
\newblock Numerical recipes : the art of scientific computing.
\newblock 3rd ed. Cambridge, UK: Cambridge University Press; 2007.
\newblock Available from: \url{https://numerical.recipes/book.html}.

\bibitem{heneghan_disentangling_2021}
Heneghan RF, Galbraith E, Blanchard JL, Harrison C, Barrier N, Bulman C, et~al.
\newblock Disentangling diverse responses to climate change among global marine ecosystem models.
\newblock Progress in Oceanography. 2021 Nov;198:102659.
\newblock Available from: \url{https://www.sciencedirect.com/science/article/pii/S0079661121001440}.

\bibitem{andersen_characteristic_2016}
Andersen KH, Berge T, Gonçalves RJ, Hartvig M, Heuschele J, Hylander S, et~al.
\newblock Characteristic {Sizes} of {Life} in the {Oceans}, from {Bacteria} to {Whales}.
\newblock Annual Review of Marine Science. 2016 Jan;8(1):217-41.
\newblock Available from: \url{https://www.annualreviews.org/doi/10.1146/annurev-marine-122414-034144}.

\bibitem{blanchard_bacteria_2017}
Blanchard JL, Heneghan RF, Everett JD, Trebilco R, Richardson AJ.
\newblock From bacteria to whales: using functional size spectra to model marine ecosystems.
\newblock Trends in ecology \& evolution. 2017;32(3):174-86.
\newblock Publisher: Elsevier.
\newblock Available from: \url{https://www.cell.com/trends/ecology-evolution/fulltext/S0169-5347(16)30236-1}.

\bibitem{nowicki_quantifying_2022}
Nowicki M, DeVries T, Siegel DA.
\newblock Quantifying the {Carbon} {Export} and {Sequestration} {Pathways} of the {Ocean}'s {Biological} {Carbon} {Pump}.
\newblock Global Biogeochemical Cycles. 2022 Mar;36(3):e2021GB007083.
\newblock Available from: \url{https://agupubs.onlinelibrary.wiley.com/doi/10.1029/2021GB007083}.

\bibitem{bianchi_estimating_2021}
Bianchi D, Carozza DA, Galbraith ED, Guiet J, DeVries T.
\newblock Estimating global biomass and biogeochemical cycling of marine fish with and without fishing.
\newblock Science Advances. 2021 Oct;7(41):eabd7554.
\newblock Available from: \url{https://www.science.org/doi/10.1126/sciadv.abd7554}.

\end{thebibliography}

\end{document}